\documentclass[namedreferences]{SolarPhysics}
\usepackage[optionalrh]{spr-sola-addons_src}
\usepackage{graphicx}
\usepackage{url}

\usepackage{times}
\newcommand{\postref}{}

\begin{document}
\begin{article}

\begin{opening}

\title{A Model for the Stray Light Contamination of the
UVCS Instrument on SOHO}

\author{S.~R.~\surname{Cranmer}$^{1}$ \sep
L.~D.~\surname{Gardner}$^{1}$ \sep
J.~L.~\surname{Kohl}$^{1}$ }

\runningauthor{S.\  Cranmer {\it et al.}}
\runningtitle{UVCS Stray Light Contamination}

\institute{$^{1}$ Harvard-Smithsonian Center for Astrophysics,
60 Garden Street, Cambridge, MA 02138, USA \\
email: \url{scranmer@cfa.harvard.edu},
\url{jkohl@cfa.harvard.edu},
\url{lgardner@cfa.harvard.edu}}

\begin{abstract}
We present a detailed model of stray-light suppression in the
spectrometer channels of the {\it Ultraviolet Coronagraph Spectrometer}
(UVCS) on the SOHO spacecraft.
The control of diffracted and scattered stray light from the bright
solar disk is one of the most important tasks of a coronagraph.
We compute the fractions of light that diffract past the UVCS
external occulter and non-specularly pass into the
spectrometer slit.
The diffracted component of the stray light depends on the finite
aperture of the primary mirror and on its figure.
The amount of non-specular scattering depends mainly on the
micro-roughness of the mirror.
For reasonable choices of these quantities, the modeled stray-light
fraction agrees well with measurements of stray light made
both in the laboratory and during the UVCS mission.
The models were constructed for the bright H~{\sc{i}} Ly$\alpha$
emission line, but they are applicable to other spectral lines
as well.
\end{abstract}

\keywords{Instrumental Effects;
Spectrum, Ultraviolet}

\end{opening}

\section{Introduction}

Until the 20th century, total solar eclipses were the only means
of observing the hot solar corona.
However, with the invention of the internally-occulted coronagraph
by Bernard Lyot in the 1930s \cite{Bi66,Ko88}, the addition of an
external occulter \cite{Ev48,NB63}, and the development of an
ultraviolet coronagraph spectrometer in the 1970s \cite{Ko78,Ko80},
a continuous and detailed exploration of coronal plasma physics
became possible.
Such instruments, combined with spectroscopic diagnostic
techniques, have become powerful tools for measuring a wide
range of plasma properties in the acceleration regions of solar
wind streams and coronal mass ejections (see, {\it e.g.,}
\opencite{Nw67}; \opencite{Wi82}; \opencite{Ko06};
\opencite{Hw08}; \opencite{Cr09}).

One of the most demanding requirements of a successful coronagraph
is the suppression of ``stray light'' scattered from the bright
solar disk.
The coronal emission tends to be many orders of magnitude less
bright than the emission from the Sun's lower atmospheric layers.
At a heliocentric distance of two solar radii ($R_{\odot}$),
the extended corona is approximately $10^{-8}$ times less bright 
than the disk at visible wavelengths.
In the ultraviolet the corona is relatively brighter, but still
is only about $10^{-6}$ (at H~{\sc{i}} Ly$\alpha$ 1216 {\AA}) to
$5 \times 10^{-6}$ (at Mg~{\sc{x}} 610 {\AA}) times the disk intensity.
These numbers correspond to the lowest-intensity regions off the
solar limb (coronal holes).
Because light rays coming from the disk are separated in angle
by only fractions of a degree from the rays coming from the
corona, there are many ways that a small fraction of the former
can easily contaminate the latter.
A key purpose of an efficient coronagraph is to suppress as much
of this stray light as possible through the use of re-imaging,
occulters, baffles, light traps, and also by making the
mirrors as smooth and accurate as possible.

This paper presents a detailed analysis of the stray light
properties of the {\it Ultraviolet Coronagraph Spectrometer}
(UVCS) instrument \cite{Ko95,Ko97} onboard the {\it Solar and
Heliospheric Observatory} (SOHO) spacecraft \cite{Dm95,FS97}.
At present, there are only a few published measurements of the
stray light properties of UVCS, and many observations have
been taken without dedicated stray-light measurements.
Thus, there is a need for a model that predicts the stray-light
intensities at ultraviolet wavelengths.
The dominant source of UVCS stray light has been shown to be
diffracted light from the external occulter that is non-specularly
scattered by the telescope mirror into the spectrometer entrance slits.
\inlinecite{Ko95} described other possible contributions to the
stray light and explained how they are controlled.
This paper primarily describes the non-specular scattering from
the telescope mirror, which accounts for essentially all of the
UVCS stray light background.

The model developed for this paper applies mainly to the Ly$\alpha$
spectrometer channel of UVCS, which observes the strong
H~{\sc{i}} Ly$\alpha$ 1216 {\AA} emission line and other spectral lines
within a wavelength range of 1100 -- 1361 {\AA}.
The model also is valid for the O~{\sc{vi}} channel, which is optimized
to observe the O~{\sc{vi}} 1032, 1037 {\AA} doublet, as well as other
spectral lines within a wavelength range of 937 -- 1126 {\AA}
in first order.
The redundant Ly$\alpha$ path of this channel can be used to
observe from 1166 to 1272 {\AA} in first order as well.
The stray-light properties of the UVCS white-light channel (WLC)
were described by \inlinecite{Ro93} and are not discussed here.

The outline of this paper is as follows.
Section 2 presents an overview of the relevant UVCS optical paths
and occulting surfaces.
Section 3 describes how much light from the solar disk is
diffracted by the UVCS external occulter, and Section 4
describes how much of this light enters the spectrometer slit
via finite-aperture diffraction and non-specular scattering.
Section 5 summarizes how the contributions from a range of locations
on the UVCS primary mirror are summed to determine the total stray
light intensity.
Section 6 presents the results of a calculation for H~{\sc{i}} Ly$\alpha$
stray light as a function of the observation height and compares
the model predictions with measurements.
Section 7 contains a brief summary of the paper and a discussion
of the capabilities of future instrumentation.
Finally, the Appendix describes how the theoretical rates of
non-specular scattering were derived separately for the effects
of mirror micro-roughness and finite-aperture diffraction.

\section{UVCS Optical Path Geometry}

The UVCS instrument contains three reflecting telescopes that
feed two ultraviolet toric-grating spectrometers and one
white-light polarimeter \cite{Ko95}.
{\postref
Unlike many other coronagraphs, the UVCS occulting surfaces are
linear, not circular, so that the incoming rays are well
matched to the linear geometry of the spectrometer entrance slit.
Rays from the bright solar disk are blocked by an external
occulter and enter a sunlight trap, whereas rays from the
extended corona are reflected by the primary mirror into the
spectrometer slits.}
These slits are oriented in the direction tangent to the solar limb.
The coronal images can be positioned on the slit centers in
heliocentric radius [$r$] anywhere between about 1.4 and 10
$R_{\odot}$ and rotated around the Sun in position angle.
The slit length projected on the sky is 40 arcminutes, or
approximately 2.5 $R_{\odot}$ in the corona, and the slit width
can be adjusted to optimize the desired spectral resolution
and count rate.

\begin{figure} 
\centerline{\includegraphics[width=0.99\textwidth,clip=]
{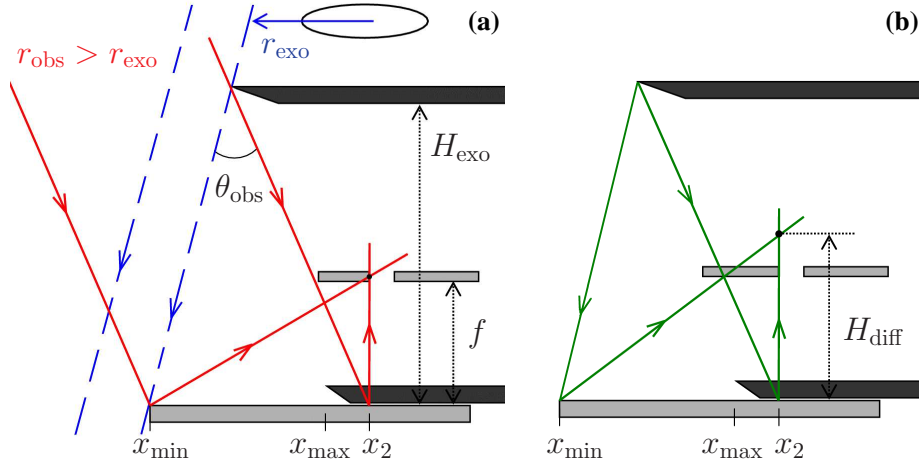}}
\caption{{\bf (a)} Schematic illustration of the UVCS/SOHO
optical path for coronal rays ({\it i.e.,} light coming from an
infinite distance).
{\bf (b)} Optical paths for rays diffracted from the tip of the
external occulter.
The horizontal distance along the mirror [$x$] increases
from left to right.
In both panels, the mirror is the light-gray surface at the
bottom, the external occulter is the dark-gray surface at the top
(at a distance $H_{\rm exo}$ above the mirror), and the
spectrometer slit is the light-gray split surface at a distance
$f$ above the mirror.
The internal occulter is the dark-gray surface sitting on top of
the mirror and covering up all $x > x_{\rm max}$.}
\label{fig1}
\end{figure}

Figure \ref{fig1}a illustrates the geometry for light coming from
the extended corona, being reflected by the UVCS primary mirror,
and entering the spectrometer slit.
The geometry is the same for the Ly$\alpha$ and O~{\sc{vi}} channels.
This diagram is not intended to be an accurate schematic of the
instrument, however.
We illustrate the mirror as being flat, but we take account
of the fact that it focuses coronal light on the slit.
The spatial scales are exaggerated in order to show how the
relevant angles and distances are defined below.
For example, all angles of reflection and scattering are small
({\it i.e.,} near normal incidence on the mirror), with
$\sin\theta \approx \tan\theta \approx \theta$ applying in
most cases.
Because the analysis of this paper is mainly concerned with
relative {\it departures} from the specularly reflected ray paths,
it is unimportant which of the rays in Figure \ref{fig1} is
portrayed as exactly normal to the mirror.

The external occulter is shown as the upper-most horizontal
surface in Figure \ref{fig1}.
It is positioned so that all specular rays from
heliocentric distances less than a given distance $r_{\rm exo}$
miss the mirror entirely and are absorbed in the UVCS sunlight
trap (see dashed blue rays).
Thus, we can only observe the extended corona at radii
$r_{\rm obs} > r_{\rm exo}$ (see solid red rays).
For UVCS, $r_{\rm exo} = 1.2 \, R_{\odot}$.
In Figure \ref{fig1}, the horizontal coordinate $x$ is measured
along the length of the mirror.
The origin $x=0$ is defined as the left edge where the ray coming
from $r_{\rm exo}$ that just barely misses the external occulter
also just barely misses the mirror.

The observation angle $\theta_{\rm obs}$, which corresponds to
the angular distance between the desired observation height and
the rays that just miss the mirror, is given by
\begin{equation}
  \theta_{\rm obs} \, = \, \theta_{\odot} \left(
  \frac{r_{\rm obs} - r_{\rm exo}}{R_{\odot}} \right)
\end{equation}
where $\theta_{\odot}$ is the angle subtended by one solar
radius on the sky.
For the mean heliocentric distance of the SOHO spacecraft,
$\theta_{\odot} = 4.698 \times 10^{-3}$ radians
$= 0.269^{\circ} = 968.9''$.
{\postref
For UVCS, the external occulter is positioned at a height of
$H_{\rm exo} = 170$ cm above the mirror.
The maximum possible region of the mirror that can be
illuminated by the corona extends from $x=0$ to
$x = H_{\rm exo} \tan\theta_{\rm obs}$.}
We can define these two limiting positions as $x_1$ and $x_2$,
respectively.
{\postref
The area of the mirror exposed to coronal light}
can be reduced by moving the UVCS internal
occulter (which sits on top of the mirror) to the left of its
default position by a distance $x_{\rm int}$.
The default (``zero over-occulting'') position corresponds exactly
to $x_2$,
{\postref
and most UVCS observations have been made with the internal
occulter positioned at $x_{\rm int} = 1.5$ mm to the left of
$x_2$.
Thus, the so-called ``vignetting function'' of UVCS is defined
by the radially increasing mirror area that is filled by coronal
rays.
For a given observation height, the unvignetted mirror area
spans the distances between $x_{\rm min} = x_{1} = 0$
and $x_{\rm max} = x_{2} - x_{\rm int}$.}

It should also be noted that neither $x_2$ nor $x_{\rm max}$ can
exceed the actual physical size of the mirror in the $x$-direction,
which for UVCS is 7.2 cm.
The corresponding maximum observation height at which the
mirror is ``filled'' is $r_{\rm obs} = 10.3 \, R_{\odot}$.

Parallel rays from an infinite distance are reflected by the mirror
and are focused at the slit, which is positioned at $f = 75$ cm
above the mirror.
For the orientation shown in Figure \ref{fig1}a, the left edge of
the slit is fixed in place (exactly at the focus point) and the
right edge of the slit is the part that opens and closes to achieve
a given slit width $w$.
In this analysis, all coronal rays -- from the coronal region of
interest -- that are reflected from mirror positions between
$x_{\rm min}$ and $x_{\rm max}$ are assumed to pass through the slit.

Figure \ref{fig1}b shows the behavior of light that is 
{\it diffracted} from the edge of the external occulter.
From the point of view of
{\postref
geometrical optics,}
these rays all appear to originate
{\postref
approximately}
at the tip of the external occulter itself ({\it i.e.,}
at a finite distance away from the mirror).
Thus, these rays are not focused to a point at $f$, but instead
are focused at a distance $H_{\rm diff}$ above the mirror, which
is given by
\begin{equation}
  \frac{1}{H_{\rm diff}} \, = \,
  \frac{1}{f} - \frac{1}{H_{\rm exo}}
  \label{eq:Hdiff}
\end{equation}
and $H_{\rm diff} = 134.21$ cm for UVCS.
Note that the specular rays in Figure \ref{fig1}b are nearly all
blocked from entering the slit. 
The only exception would be the right-most rays that strike the
mirror at $x_{2} = H_{\rm exo} \tan\theta_{\rm obs}$.
These are eliminated by just a small amount of over-occulting
by the internal occulter.
However, the diffracted beams also have substantial energy at
non-specular angles, and these are discussed further in Section 4.

\section{Diffraction from the External Occulter}

The external occulter is the first line of defense against
stray-light contamination.
In this section, we compute how much light is diffracted by
the external occulter and falls onto the mirror.
We ignore the diffracted light that does not hit the
mirror, but instead strikes other structural components of the
instrument where it can undergo multiple reflections before
possibly finding its way to the entrance slit.
Specifically, we ignore the stray light produced by scattering
off the mirror edge and the surface of the internal occulter.
These effects can be important at visible wavelengths, but they
have been shown to be negligible in the ultraviolet for
externally-occulted designs like UVCS (see, {\it{e.g.,}}
\opencite{Ro93}; \opencite{Ko95}).

The problem of parallel rays diffracting around an ideal
one-dimensional straight edge ({\it i.e.,} a semi-infinite screen) is
treated in standard textbooks ({\it e.g.,} \opencite{BW99};
\opencite{Hecht2002}).
In practice, it has been found that serrated, or ``toothed'' edges
perform better than ideal straight edges \cite{NB63,KB87,Vr08}.
In the late 1980s, laboratory measurements were made at the
Smithsonian Astrophysical Observatory for both straight-edged
and serrated occulters, in combination with the {\it Spartan 201}
UV coronagraph spectrometer (see \opencite{Ro93}).
The laboratory measurements were done with a light source that
subtends the same solid angle as the solar disk.
The serrated occulter used on UVCS has symmetric sawtooth-shaped
notches ({\it i.e.,} resembling a repeating triangle waveform)
with a separation of the peaks, parallel to the occulter edge, of
229 $\mu$m.
The depth of the peaks, perpendicular to the occulter edge, is
267 $\mu$m.

\begin{figure} 
\centerline{\includegraphics[width=0.90\textwidth,clip=]
{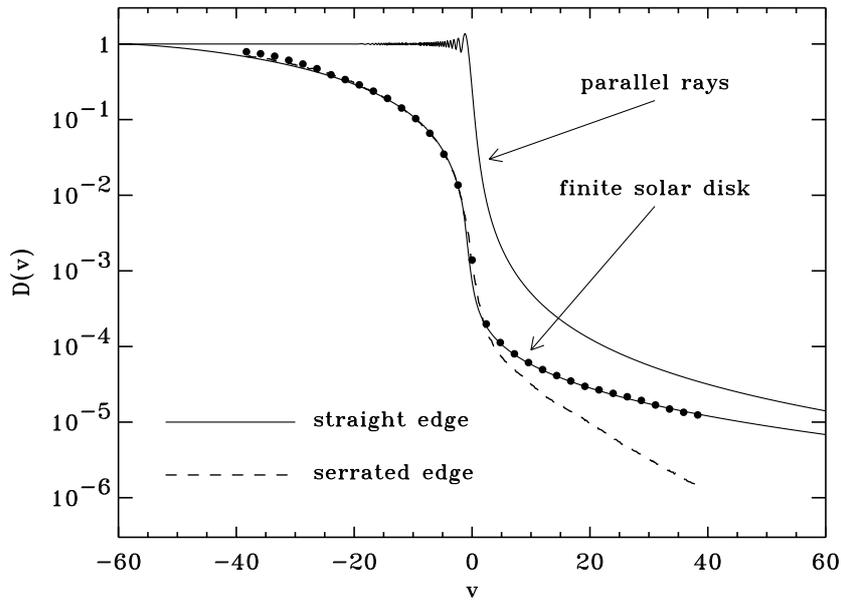}}
\caption{Fractional reduction in irradiance due to diffraction at the
external occulter, shown as a function of dimensionless distance
[$v$] behind the occulter.  Laboratory measurements for a straight
edge (points) and a serrated edge (dashed curve) are compared with
computed diffraction curves (solid curves).}
\label{fig2}
\end{figure}

Figure \ref{fig2} illustrates laboratory measurements and
theoretical predictions for the reduction in irradiance
({\it{i.e.,}} radiative flux) seen as an observer moves
progressively into the occulter's shadow.
This quantity is shown specifically for the H~{\sc{i}} Ly$\alpha$
wavelength, and the distance behind the occulter was converted
into dimensionless Fresnel diffraction coordinates
\begin{equation} 
  v \, = \, (x - x_{\odot}) \sqrt{\frac{2}{\lambda H_{\rm exo}}}
  \label{eq:vdef}
\end{equation}
where $\lambda$ is the wavelength of interest and $x_{\odot}$ is
the mirror position that corresponds to rays from the near edge of
the solar disk that barely miss the external occulter.
Because these rays do not strike the mirror, $x_{\odot}$ is negative
in our coordinate system, and it is given by
$-H_{\rm exo} \tan (\theta_{\odot} [r_{\rm exo}/R_{\odot} - 1])$.
Thus, all values of $v$ that are relevant for UVCS are positive.
We denote the relative irradiance quantity plotted in
Figure \ref{fig2} as $D(v)$.

The uppermost curve in Figure \ref{fig2} corresponds to
parallel rays ({\it i.e.,} from a point-source on the sky at
infinite distance).  This is given by
\begin{equation}
  D(v) \, = \, \frac{1}{2} \left\{
  \left[ \frac{1}{2} - C(v) \right]^{2} +
  \left[ \frac{1}{2} - S(v) \right]^{2} \right\}
\end{equation}
where $C(v)$ and $S(v)$ are the standard Fresnel integrals
\cite{AS72}.
When observing the Sun, this function must be convolved with the
finite size of the solar disk on the sky, since the total diffracted
light is a combination of beams between the ``near edge'' of the
solar disk ({\it i.e.,} only 0.2 $R_{\odot}$ behind the $r_{\rm exo}$
ray that just grazes the left edge of the mirror) and the
``far edge'' (up to 2.2 $R_{\odot}$ behind the grazing ray).
{\postref
For the H~{\sc{i}} Ly$\alpha$ emission line, we consider the solar disk
to be uniformly bright \cite{Cu08}.}
The result of this numerical convolution (lower solid curve)
agrees very well with the measured data for a straight-edge
occulter (points) taken with a light source with the same finite
solid angle as the Sun (see also \opencite{Ro93}).

It can also be seen from Figure \ref{fig2} that the serrated
edge produced a significantly lower amount of diffraction than
the straight edge.
The dashed curve corresponding to the serrated edge is denoted
$D_{s}(v)$.
The laboratory measurements extended up to an occulting distance
corresponding to $v \approx 38$.
UVCS observations made at $r_{\rm obs} \leq 2.74 \, R_{\odot}$
fall into the range of heights where the laboratory measurements
can be applied via interpolation.
However, UVCS observations at larger heights require some degree
of extrapolation beyond the measured range.
For example, the largest height of
$r_{\rm obs} \approx 10 \, R_{\odot}$ fills the mirror with a
range of $v$ values between about 5 and 220.
After some experimentation, we found that a power-law extrapolation
curve that scales as $v^{-3}$ best reproduced and continued
the slope of the measured data.
\inlinecite{Ro93} used an exponential function which may become
lower than the $v^{-3}$ curve at large values of $v$.
Thus, our power-law extrapolation may give rise to a slight (but
conservative) overestimate of the diffraction at large observation
heights.

\section{Non-specular Rays Entering the Slit}

\begin{figure} 
\centerline{\includegraphics[width=0.70\textwidth,clip=]
{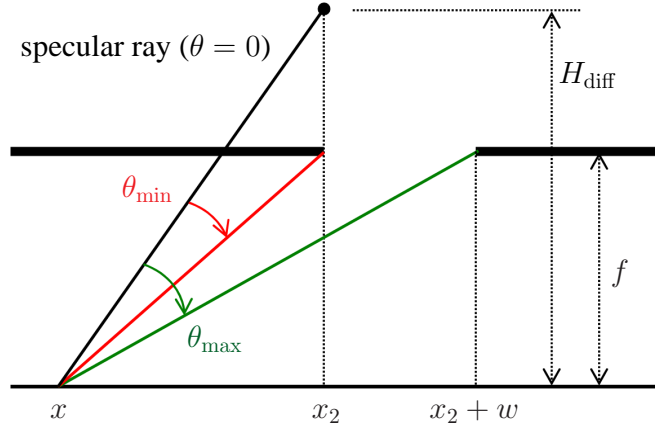}}
\caption{Non-specular geometry for diffracted rays originating at
an arbitrary point [$x$] on the mirror surface and passing through
the spectrometer slit.}
\label{fig3}
\end{figure}

Each point on the mirror between $x_{\rm min}$ and
$x_{\rm max}$ acts as a ``source'' of solar-disk light that
has been diffracted by the external occulter.
These specular rays are focused at a point $H_{\rm diff}$ above
the mirror, and are blocked by the left-side slit edge.
However, the light is not concentrated all along the ideal
specular ray, but instead has a characteristic {\it non-specular}
spread that subtends a broader range of solid angle.
Thus, there are always non-specular rays from the mirror
that are able to pass through the slit.
Figure \ref{fig3} illustrates those rays, which are denoted by
non-specular angles ranging between $\theta_{\rm min}$ and
$\theta_{\rm max}$.
For a given point [$x$], the values of these angles are given by
\begin{equation}
  \theta_{\rm min} \, = \,
  \tan^{-1} \left( \frac{H_{\rm diff}}{x_{2} - x} \right) -
  \tan^{-1} \left( \frac{f}{x_{2} - x} \right)
  \label{eq:tmin}
\end{equation}
\begin{equation}
  \theta_{\rm max} \, = \,
  \tan^{-1} \left( \frac{H_{\rm diff}}{x_{2} - x} \right) -
  \tan^{-1} \left( \frac{f}{x_{2} + w - x} \right)
  \label{eq:tmax}
\end{equation}
where $w$ is the slit width. 
Note that for the special case of $x = x_{2}$, some specular
rays pass through the slit since $\theta_{\rm min} = 0$ and
$\theta_{\rm max} = \tan^{-1} (w/f)$.
Of course, for any finite amount of internal over-occulting,
these rays never get reflected at all, since we endeavor to keep
all rays that reach the mirror at $x < x_{\rm max} < x_{2}$.

The fraction of reflected light that passes non-specularly into
the spectrometer slit is computed by summing the effects of two
distinct physical processes:
{\it{i}}) diffraction due to the finite size of the mirror, and
{\it{ii}}) scattering due to mirror micro-roughness.
The Appendix gives a derivation of these two components.
{\postref
We note that this separation into two parts is an approximation
of the full angular distribution of non-specular intensity,
which for UVCS is not known with sufficient detail to specify
as a single function.
It is useful to specify these two effects separately so that
different assumptions about their origin and magnitude can be
made without each ``contaminating'' the other.
Taking their direct sum may result in an amount of non-specular
scattering that is slightly larger than the actual combined effect
of these two processes, but this errs on the side of making a
conservative overestimate of the stray light.
The combined differential scattering and diffraction can thus be
expressed as}
\begin{equation}
  \frac{\mathrm{d}I}{I_{0} \mathrm{d} \theta_x} \, = \,
  \frac{\lambda}{2 \pi^{2} D_{x} \theta_{x}^{2}} \left[
  W + \frac{D_x}{\Lambda} \left(
  \frac{32 \pi^{3} \sigma^2}{\lambda^2} \right) \right] \,\, ,
  \label{eq:Sboth}
\end{equation}
where $D_{x} = x_{\rm max} - x_{\rm min}$ is the illuminated
mirror width, $\theta_x$ is the non-specular angle measured in
the plane of Figure \ref{fig3}, $W$ is a wing enhancement factor
that takes account of deviations from a perfect diffraction
aperture, $\sigma$ is the root-mean-squared (r.m.s.) micro-roughness
of the mirror, and $\Lambda$ is a horizontal coherence length for
the mirror's non-ideal surface modulations.
The first term in square brackets above takes account of
diffraction due to the mirror's finite size, and the second term
accounts for micro-roughness.

Equation (\ref{eq:Sboth}) assumes both that the micro-roughness is
small compared to the wavelength ({\it i.e.,} $\sigma \ll \lambda$)
and that the non-specular angles are large compared to the angular
size of the specular beam ({\it i.e.,}
$\theta_{x} \gg \lambda / D_{x}$).
We denote the relative fraction of the diffracted solar-disk
light on the mirror surface that makes it through the slit
as $\mathrm{d}I / I_{0}$.
This fraction is given by integrating over all non-specular
angles between $\theta_{\rm min}$ and $\theta_{\rm max}$, and
\begin{equation}
  \frac{\mathrm{d}I}{I_0} \, = \,
  \int_{\theta_{\rm min}}^{\theta_{\rm max}}
  \mathrm{d}\theta_{x} \, \frac{\mathrm{d}I}
  {I_{0} \mathrm{d}\theta_x} \,\, .
\end{equation} 
{\postref
Note that the integration over the other non-specular angle
$\theta_y$ is discussed in the Appendix.}

\section{Flux Normalization}

Next, it is necessary to compute the total flux that comes from
a given location on the mirror and passes through a
given part of the slit.
We must take account of the full two-dimensional shape of the slit
aperture; {\it i.e.,} not just the width [$w$] but also the height
[$h$] of the spatial element of interest (in the
direction normal to the page in Figure \ref{fig1}).
From the point of view of the mirror, the solid angle of the slit
is given by $\Omega_{\rm slit} = wh / f^{2}$.
The flux of light from the extended corona is thus given by
\begin{equation}
  F_{\rm cor}(x) \, = \, \left\{
  \begin{array}{cc}
    I_{\rm cor} \Omega_{\rm slit} \,\, , &
    x_{\rm min} < x < x_{\rm max} \\
    0 & \mbox{otherwise}
  \end{array} \right.
  \label{eq:Fcor}
\end{equation}
where $I_{\rm cor}$ is the coronal intensity, in units of
erg s$^{-1}$ cm$^{-2}$ sr$^{-1}$ ({\it i.e.,} integrated across
the spectral line).
We assume that $I_{\rm cor}$ is constant over the slit aperture.

The stray light, which began its life as the direct solar-disk
intensity $I_{\odot}$, is attenuated by diffraction and
non-specular scattering.
Without this attenuation, the flux from the solar disk is
given by $I_{\odot} \Omega_{\odot}$, where $\Omega_{\odot}$
is the solid angle of the Sun in the sky (essentially
$\pi \theta_{\odot}^{2}$).
However, because of the finite slit height $h$, only a fraction
of this flux would make it through the slit.
In the limiting case where the angle subtended by the slit-height
direction (approximately $h/f$) is small compared to the solar
diameter ($\delta_{\odot} = 2 \theta_{\odot}$), the piece of the
solar disk that is ``seen'' through the slit is only a strip
with solid angle $\Omega_{\rm strip} \approx h \delta_{\odot}/f$.
In the opposite limit of $h/f \geq \delta_{\odot}$, then
$\Omega_{\rm strip} = \Omega_{\odot}$.
The code used to compute the stray light uses an exact
expression that bridges these limiting cases, which for
$h/f \leq \delta_{\odot}$ is given by
\begin{equation}
  \Omega_{\rm strip} \, = \, \frac{\delta_{\odot}^2}{2} \tan^{-1}
  \left( \frac{h}{\sqrt{\delta_{\odot}^{2} f^{2} - h^2}} \right)
  + \frac{h \sqrt{\delta_{\odot}^{2} f^{2} - h^2}}{2f^{2}}
  \,\, .
\end{equation}
For all of the numerical calculations shown below, however, the
approximation $\Omega_{\rm strip} \approx h \delta_{\odot}/f$ is
reasonably valid.
Thus, the fully attenuated stray-light flux is
\begin{equation}
  F_{\rm stray}(x) \, = \, I_{\odot} \Omega_{\rm strip} \,\,
  D_{s} (v) \, \frac{\mathrm{d}I}{I_0} \,\, .
\end{equation}

The two fluxes defined above represent power per unit mirror
surface area.
The total power, in erg s$^{-1}$, that passes through the slit
for the two cases is given by integrating over the mirror surface:
\begin{equation}
  P_{\rm cor} \, = \, \int\int \mathrm{d}x \, \mathrm{d}y
  \,\, F_{\rm cor}(x)
\end{equation}
\begin{equation}
  P_{\rm stray} \, = \, \int\int \mathrm{d}x \, \mathrm{d}y
  \,\, F_{\rm stray}(x)
\end{equation}
where the integration over the $y$-coordinate (the mirror dimension
out of the page in Figure \ref{fig1}) is trivial because the
fluxes are independent of $y$.
The $x$-coordinate ranges from $x_{\rm min}$ to $x_{\rm max}$.
The $y$-coordinate ranges from $y_{\rm min} = 0$ to
$y_{\rm max} = 5$~cm (the latter being the UVCS mirror height).

Finally, the ``calibrated'' measurement for stray-light intensity
can be constructed.
Once the total power quantities defined above are known, the
corresponding intensities can be computed simply by dividing either
$P_{\rm cor}$ or $P_{\rm stray}$ by the product
$A \Omega_{\rm slit}$.
The illuminated mirror area is given by
\begin{equation}
  A \, = \, (y_{\rm max} - y_{\rm min})
  (x_{\rm max} - x_{\rm min})  \,\,\, .
\end{equation}
This process is unnecessary for the coronal intensity, since
$I_{\rm cor}$ was used in its initial definition
[Equation (\ref{eq:Fcor})].
However, the stray-light intensity [$I_{\rm stray}$] is computed in
this way by dividing $P_{\rm stray}$ by
$A \Omega_{\rm slit}$ directly.
{\postref
We do not consider any instrumental properties beyond the entrance
slit ({\it e.g.,} efficiencies of the grating and detector) because
they do not distinguish between coronal and stray light.}

\section{Results}

We implemented the above constraints on ray geometry,
external-occulter diffraction, non-specular scattering, and
flux calibration into an IDL code that predicts the stray-light
intensity [$I_{\rm stray}$] for various configurations of the
UVCS instrument.
A standard set of models for the H~{\sc{i}} Ly$\alpha$ spectral line
was constructed with the geometric parameters given in the
previous sections, as well as the following choices for the
internal occulter position ($x_{\rm int} = 1.5$ mm),
slit width ($w = 100$ $\mu$m), and the angular resolution of
the spatial element of interest along the slit (10 arcseconds,
or $h = 36.4$ $\mu$m on the slit).

{\postref
In the models presented here, the}
two main properties of the mirror that describe its
diffraction and non-specular scattering are the
r.m.s.\  micro-roughness [$\sigma$] and the
r.m.s.\  wavefront error [$\omega$].
As described in the Appendix, the wavefront error is a single
parameter that characterizes the large-scale figure error of
the mirror.
This parameter determines the angular width of the specular beam
and the enhancement of the large-angle diffraction wing.
The figure error can also be expressed as an effective mirror
size $D_{\rm eff}$, which acts as a large-scale coherence length
analogous to the micro-roughness coherence length [$\Lambda$].
Although no precise measurements of these quantities exist for
the UVCS mirrors, there are several reasonable lower and upper
bounds that can be applied.

First, a ``best-case'' model can be constructed with the
assumption of zero micro-roughness.
For $\sigma = 0$, the non-specular rays are solely the result of
diffraction around the finite aperture of the primary mirror.
The lowest possible level of non-specular intensity is consistent
with the diffraction-limited case of no figure error ($\omega = 0$).
This limit is interesting because it provides a true lower bound
on the stray light that cannot be improved upon without changing
the instrument design.

Second, we constructed a model with empirical upper limits on
the micro-roughness and figure error.
{\postref
Scattering measurements were made with a helium-neon laser on the
mirror eventually chosen for flight as the Ly$\alpha$ channel
primary \cite{Lv98}.
These measurements were consistent with a value for the
r.m.s.\  micro-roughness of approximately 20 {\AA}.
We note that this determination is dependent on a particular
numerical model for the scattering properties of mirrors, which has
been described by \inlinecite{Sh96}.
Nonetheless, we use $\sigma \approx 20$ {\AA} as an approximate
upper limit for the UVCS primary mirrors.}
Two independent determinations of the spatial coherence lengths of
the mirror were found that agree with one another reasonably well:
\newcounter{bean}
\begin{list}{\arabic{bean}.}{\usecounter{bean}%
\setlength{\leftmargin}{0.18in}%
\setlength{\rightmargin}{0.0in}%
\setlength{\labelwidth}{0.18in}%
\setlength{\labelsep}{0.05in}%
\setlength{\listparindent}{0.0in}%
\setlength{\itemsep}{0.05in}%
\setlength{\parsep}{0.0in}%
\setlength{\topsep}{0.05in}}
\item[{\it{i}})]
We used the \inlinecite{Lv98} measurements of non-specular
scattering to put a limit on the mirror's coherence length.
The large-angle limit of Equation (\ref{eq:Sinha}) can be
used to show that
\begin{equation}
  \Lambda \, \approx \, \frac{8\pi \sigma^2}{\lambda \theta^3}
  \left( \frac{\mathrm{d}I}{I_{0} \mathrm{d}\Omega} \right)^{-1} \,\, .
  \label{eq:LamLev}
\end{equation}
Measurements of $\mathrm{d}I / I_{0} \mathrm{d}\Omega$ were taken
at angles [$\theta$] corresponding to heliocentric distances between
0.95 and 11.3 $R_{\odot}$.
The mean value of $\Lambda$ for all of these measurements,
derived from Equation (\ref{eq:LamLev}), was found to be 1.36 cm.
This results in a ratio $[D_{x} / \Lambda] = 5.3$ for the
full unvignetted mirror width $D_{x} = 7.2$ cm.
\item[{\it{ii}})]
Another determination of the mirror's performance can be
found from the measured UVCS spatial resolution.
The most definitive upper limits on spatial resolution were
obtained from in-flight observations of the star $\rho$ Leo,
made in August 1996.
For H~{\sc{i}} Ly$\alpha$, the measured full-width at half-maximum
(FWHM) of the spatial resolution element was found to be 5.7 arcseconds.
Although some fraction of this value may be attributable to
other instrumental factors, it is useful to assign it all to
the mirror's figure error to obtain an upper limit.
Ideally, the FWHM should be converted into an angular half-width
[$\theta_s$] as defined in the Appendix.
If the measured profile were a Gaussian beam, it would encompass
83.8\% of the encircled intensity out to an angular half-width of
about 0.8 times the FWHM.
Because this is so close to unity, and because its exact shape is
unknown, we just use the FWHM itself to assume an upper-limit
value of $\theta_{s} \approx 5.7$ arcseconds.  Thus,
Equation (\ref{eq:thetas}) gives $D_{\rm eff} \approx 0.54$ cm.
The stellar observations were taken at a heliocentric distance
$r_{\rm obs} \approx 5.2 \, R_{\odot}$ at which the
unvignetted mirror width was $D_{x} \approx 3.03$ cm.
The derived ratio $D_{x} / D_{\rm eff}$ was found to be 5.6.
\end{list}
Because the ratios $D_{x} / \Lambda$ and $D_{x} / D_{\rm eff}$
were found to be in good agreement with one another, we
make the tentative assumption that one can treat
$\Lambda$ and $D_{\rm eff}$ roughly interchangeably for
this analysis.
Thus, we adopt a mean ratio $D_{x} / \Lambda = 5.45$ and we
use Figure \ref{fig5} to give the corresponding
wavefront error $\omega / \lambda \approx 0.23$ and wing
enhancement ratio $W \approx 1.78$.

\begin{figure} 
\centerline{\includegraphics[width=0.93\textwidth,clip=]
{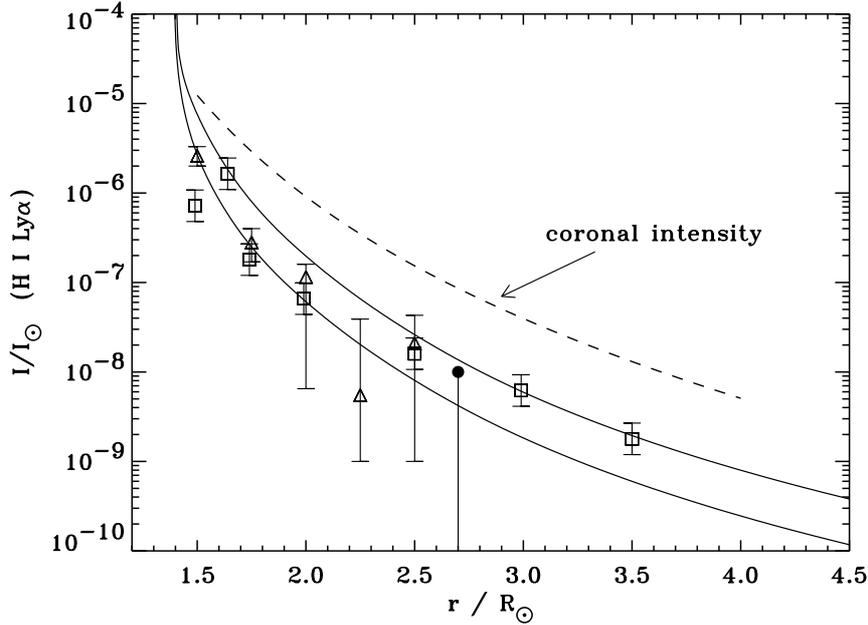}}
\caption{Comparison of modeled and observed stray-light
suppression ratios $I_{\rm stray} / I_{\odot}$.
Models are shown for the lower-limit case of no micro-roughness
and diffraction-limited optics (lower solid curve) and
an empirical upper limit on the mirror parameters
(upper solid curve).
The stray-light measurements (points) are described in the text.
For comparison, the H~{\sc{i}} Ly$\alpha$ intensity measured by
UVCS in a polar coronal hole is also shown (dashed curve).}
\label{fig4}
\end{figure}

Figure \ref{fig4} shows the numerically computed
$I_{\rm stray} / I_{\odot}$ ratios that correspond to the
two limiting sets of parameters described above.
The lower-limit set of mirror parameters ($\sigma = 0$,
$D_{x} / \Lambda = 1$) shows the effect of having only
diffraction contribute to the stray light.
For the empirical upper-limit parameters ($\sigma = 20$ {\AA},
$D_{x} / \Lambda = 5.45$) the micro-roughness contributed
roughly half (45\%) of the total stray light.
The following parameterized fits were found to be good
representations of the model curves in Figure \ref{fig4},
\begin{equation}
  \left( \frac{I_{\rm stray}}{I_{\odot}} \right)_{\rm lower}
  \, = \,
  \frac{3.14 \times 10^{-8}}{([r / R_{\odot}] - 1.121)^{4.56}}
  \,\, ,
\end{equation}
\begin{equation}
  \left( \frac{I_{\rm stray}}{I_{\odot}} \right)_{\rm upper}
  \, = \,
  \frac{1.30 \times 10^{-7}}{([r / R_{\odot}] - 1.076)^{4.73}}
  \,\, ,
\end{equation}
and these fits apply mainly to heights between 1.45 and 5
$R_{\odot}$.

The measured stray-light data points in Figure \ref{fig4}
were obtained from several sources.
Pre-launch laboratory testing provided the upper limit value
of $10^{-8}$ (filled circle) at $r = 2.7 \, R_{\odot}$ \cite{Ko95}.
Observations of the faint Si~{\sc{iii}} 1206 {\AA} emission
line -- which has no coronal counterpart and is 100\% stray light
when measured by UVCS -- were made in June 1996 (triangles) and
scaled to the disk intensity of H~{\sc{i}} Ly$\alpha$ to obtain the
stray-light suppression ratio \cite{Ga96}.
Additional data points obtained from measurements of the
C~{\sc{iii}} 977 {\AA} line (squares) are also shown.
Several of these measurements were discussed by \inlinecite{Ko99}.
The remaining measurements are those at 1.64 $R_{\odot}$
\cite{Ry03} and at 3.5 $R_{\odot}$ \cite{Su99}.
In the latter two cases, only the total intensities were
reported in the cited papers.
These values were normalized to the solar disk and scaled to
H~{\sc{i}} Ly$\alpha$ using the relative line intensities of
\inlinecite{VR78} and the time-dependent irradiance measurements
of \inlinecite{Wo00}.
The error bars on the C~{\sc{iii}} points were estimated from the
assumption of an overall 50\% uncertainty, which can be split
up evenly between 25\% uncertainties in the numerator
[$I_{\rm stray}$] and the denominator [$I_{\odot}$].

Figure \ref{fig4} does not show data for distances larger than
$r_{\rm obs} \approx 4 \, R_{\odot}$.
At these large observation heights, the measured stray light is
dominated by a roughly constant background.
This background is likely to be the result of several effects,
including the dust-scattered F-corona, interstellar line emission
(for H~{\sc{i}} Ly$\alpha$ and other lines of neutral species), and
possibly other sources of instrumental stray light that are
negligible at lower heights.
{\postref
In visible-light Lyot coronagraphs, a constant stray-light
background of this kind is often the result of dust particles on
the primary mirror \cite{Ne08}.
However, for the externally-occulted design of UVCS -- in which
the relatively small primary mirror is never exposed to direct
sunlight -- the effects of dust have been considered negligible at
both visible and ultraviolet wavelengths \cite{Ro93,Ko95}.}
The vast majority of UVCS coronal measurements have been made
within the range of heights shown in Figure \ref{fig4}.

For any given observation height, the intensity of the
extended corona can vary by several orders of magnitude.
Figure \ref{fig4} shows the mean radial dependence of the
intensity of H~{\sc{i}} Ly$\alpha$ in polar coronal holes measured
during the 1996 -- 1997 minimum of solar activity \cite{Cr99}.
Coronal holes are the lowest-intensity structures in the extended
corona, and thus they represent the most stringent requirements
on stray-light suppression.
The coronal intensities were normalized by dividing by a disk
intensity typical for the time period of these solar-minimum
observations, $I_{\odot} = 5.3 \times 10^{15}$ photons s$^{-1}$
cm$^{-2}$ sr$^{-1}$ ({\it{e.g.,}} \opencite{Wo00}).
Even the emission from coronal holes is significantly larger
than the modeled and measured stray-light levels for
H~{\sc{i}} Ly$\alpha$.
Other spectral lines tend to exhibit larger ratios of
$I_{\rm cor} / I_{\odot}$ than H~{\sc{i}} Ly$\alpha$, so the
stray light is less of a contaminant.

In order to scale the above results to other spectral lines,
the overall wavelength dependence of $I_{\rm stray}$ must
be calculated.
The diffraction from the external occulter depends on $\lambda$
because of the definition of $v$ in Equation (\ref{eq:vdef}).
Also, the total rate of diffraction and scattering through the
slit [Equation (\ref{eq:Sboth})] scales differently with $\lambda$
depending on whether the non-specular stray light is dominated
by finite mirror-size diffraction or by micro-roughness.
At a standard observation height of 2 $R_{\odot}$, we computed
$I_{\rm stray}$ for a range of wavelengths between half and twice
that of H~{\sc{i}} Ly$\alpha$.
The lower-limit case of $\sigma = \omega = 0$ gives rise to an
approximate wavelength dependence
$I_{\rm stray} \propto \lambda^{1.91}$.
The empirical upper-limit choice of mirror parameters gives
$I_{\rm stray} \propto \lambda^{0.74}$.
These exponents increase slowly as a function of increasing height.
For the bright O~{\sc{vi}} 1032, 1037 {\AA} doublet, there is only a
15\% relative separation in wavelength from 1216 {\AA}.
Since the uncertainties in the model parameters exceed this level
at any given height, it is reasonable to just use the modeled
value of $I_{\rm stray}$ that was computed for H~{\sc{i}} Ly$\alpha$.

\section{Conclusions}

The primary aim of this paper has been to construct a model of
the ultraviolet stray-light properties of the UVCS instrument
on SOHO.
The model has essentially no freely adjustable parameters, with
the possible exception of the parameters $\sigma$ and $\omega$
that define the mirror imperfections.
Estimates of these parameters were obtained from pre-launch
laboratory tests and in-flight measurements.
{\postref
A completely separate set of existing measurements of the
stray-light suppression ratio for H~{\sc{i}} Ly$\alpha$ was used to
test the model.
As can be seen in Figure \ref{fig4}, the modeled stray light
agrees well with these measurements.}

It is interesting to investigate how much of an improvement in
stray-light suppression can be made by various changes to the
externally occulted coronagraph design exemplified by UVCS.
Equation (\ref{eq:Sboth}) indicates that even if the mirror
were made to be perfectly smooth ($\sigma = 0$) and ideally
diffraction-limited ($W=1$, $D_{x} = \Lambda$), there would
still be a finite level of non-specular stray light due to
diffraction by the finite-sized mirror.
Reductions in the amount of non-specular radiation can then
be achieved by increasing the mirror size [$D_x$] or by
increasing the magnitudes of the non-specular angles
[$\theta_x$] that pass through the slit.

Designs for next-generation coronagraph spectrometers have also
achieved improvements in performance by increasing the distance
[$H_{\rm exo}$] between the mirror and external occulter
\cite{Ko06,Ko08}.
An increase in $H_{\rm exo}$ results in a smaller value of
$H_{\rm diff}$ [Equation (\ref{eq:Hdiff})], which leads to smaller
non-specular angles [$\theta_x$].
Taken by itself, this would give rise to an increase in the
non-specular intensity that scatters and diffracts from the
mirror into the slit.
However, this effect is more than offset by the larger
unvignetted mirror area that becomes available when $H_{\rm exo}$
is increased ({\it i.e.,} an increase in $x_2$).
The larger mirror area gives rise to a ``deeper shadow'' under
the external occulter, which is equivalent to a larger extent of
$v$ in Figure \ref{fig2}.
This results in a lower overall amount of diffracted light
striking the mirror.
Also, the larger available mirror area allows more coronal rays
to be collected, thus increasing the overall instrumental
sensitivity.
These advantages, together with improvements in mirror
reflectivity and detector efficiency, can give rise to
several orders of magnitude improvement in sensitivity
compared to UVCS/SOHO.

\appendix 

\section{Derivation of Non-Specular Intensity Profiles}

Here we describe the angular dependence of non-specular
diffraction and scattering that arise from reflection (at
nearly normal incidence) from a mirror of finite size and with
a non-ideal surface.
The fraction of the incident intensity $I_0$ scattered into a
differential solid angle [$\mathrm{d}\Omega$] is defined here as
\begin{equation}
  S ({\bf q}) \, = \, \frac{\mathrm{d}I}{I_{0} \mathrm{d}\Omega}
  \,\, , \label{eq:dIdef}
\end{equation}
where the scattering wavevector ${\bf q}$ is the vector difference
between the incoming and reflected photon wavevectors ${\bf k}$.
Thus, if the mirror surface is assumed to lie in the $x,y$
plane, then $(q_{x},q_{y},q_{z}) \approx 2\pi
(\theta_{x}, \theta_{y}, 2)/\lambda$.
The goal of this Appendix is to show how $S$ depends explicitly on
the non-specular angles $\theta_{x}$ and $\theta_{y}$.
The specularly reflected ray is defined as $\theta_{x} = \theta_{y} = 0$.

{\postref
In the optics literature, the differential scattering fraction
defined above is often described using other terms.
A frequently used variant is the bi-directional reflectance
distribution function (BRDF), which is defined as
\begin{equation}
  \mbox{BRDF} \, = \, \frac{1}{\cos \Theta}
  \frac{\mathrm{d}I}{I_{0} \mathrm{d}\Omega}  \,\, ,
\end{equation}
and where $\Theta$ is the scattering angle measured from the
local normal to the mirror surface \cite{Bn99}.
Another commonly used term is the power spectral density (PSD),
which is a measure of the angular distribution of fluctuations
on the mirror surface itself.
The PSD can be expressed in similar units as the
angular distribution of reflected power, and
\begin{equation}
  \mbox{PSD} \, = \, \frac{\lambda^{4}}{16 \pi^{2} {\cal R}}
  \frac{\mathrm{d}I}{I_{0} \mathrm{d}\Omega}  \,\, .
\end{equation}
The dimensionless quantity ${\cal R}$ is the overall reflectivity
of the mirror, and it is often called an ``optical factor'' that
depends on wavelength, the angles of incidence and reflection,
and the degree of polarization of the incident and reflected
beams \cite{Bn03}.
In order to separate out the problem of mirror reflectivity from
the problems of non-specular scattering and diffraction, the
analysis below makes the assumption that ${\cal R} = 1$.}

The angular intensity profile can be derived by applying the Born
approximation from scattering theory ({\it e.g.,}
\opencite{Si88}; \opencite{CT93}; \opencite{Gu97}).
If the reflectivity of the mirror is assumed to be perfect, the
Born approximation can be expressed as
\begin{equation}
  S({\bf q}) \, = \, \frac{1}{A \lambda^{2}} \int
  \mathrm{d}x \, \mathrm{d}y \int
  \mathrm{d}x' \, \mathrm{d}y' 
  \, e^{-i {\bf q} \cdot ({\bf r} - {\bf r}')}
  \,\, .  \label{eq:born}
\end{equation}
Both sets of spatial integrals are taken over the mirror area $A$
in the $x,y$ plane.
The position vector ${\bf r}$ depends not only on the horizontal
position, but also on the spatially varying mirror height $z(x,y)$.

In the UVCS instrument, the most important orientation for the
non-specular diffraction and scattering is the $\theta_x$
direction ({\it i.e.,} the left--right horizontal direction in
Figures \ref{fig1} and \ref{fig3}).
Because light from a given point in the $y$-direction (on the
sky or on the external occulter) fills the mirror in the
$y$-direction, there ends up being a wide range of $\theta_y$
scattering angles that go through the slit.
Thus, we first solve Equation (\ref{eq:born}) and then
integrate over all $\theta_y$ angles to obtain the differential
scattering rate $\mathrm{d}I / (I_{0} \mathrm{d}\theta_{x})$.
Sections A.1 and A.2 give independent calculations of this
quantity in two limiting regimes of the spatial scales
of mirror imperfections.
Equation (\ref{eq:Sboth}) above estimates their combined effect
as the sum of the two rates.

\subsection{Non-Specular Scattering due to Micro-Roughness}

When only small-scale statistical fluctuations in the mirror surface
are taken into account, it is possible to describe the mirror height
with spatially averaged moments.
The mean height is defined as
\begin{equation}
 \langle z \rangle \, = \, \frac{1}{A} \int
 \mathrm{d}x \, \mathrm{d}y \, z(x,y)
  \, = \, 0 \,\, ,
\end{equation}
and the r.m.s.\  surface roughness is given by
$\langle z^{2} \rangle^{1/2} = \sigma$.
{\postref
As defined here, $\sigma$ contains no information about the
horizontal distribution of microscopic structures on the mirror
surface.
Typically, laboratory measurements of $\sigma$ are dependent on
the range of horizontal scales that were probed.
Many real surfaces have a ``fractal'' distribution of surface
fluctuations that are ill-described by a single $\sigma$.
To begin the process of characterizing the horizontal scales,
though,}
we define an autocorrelation function
\begin{equation}
  C(s,t) \, = \, \frac{1}{A} \int
  \mathrm{d}x \, \mathrm{d}y \,\, z(x,y) \, z(x-s, y-t)
\end{equation}
and we assume the surface is isotropic on small enough scales
such that $C$ is a function of only $r \equiv (s^{2} + t^{2})^{1/2}$.

With the above definitions, the scattering integrals in
Equation (\ref{eq:born}) can be shown to reduce to
\begin{equation}
  S(\theta) \, = \, \frac{2\pi}{\lambda^2} \int
  \mathrm{d}r \, r \,
  \exp \left\{ - \frac{16 \pi^2}{\lambda^2} \left[ \sigma^{2}
  - C(r) \right] \right\} J_{0} \left( \frac{2\pi r \theta}{\lambda}
  \right) 
\end{equation}
where $J_0$ is the zeroth-order Bessel function of the first kind,
and $\theta = (\theta_{x}^{2} + \theta_{y}^{2})^{1/2}$ is the
full magnitude of the scattering angle away from the specular beam.
In this section we assume the integration over $r$ can be taken
from zero to infinity, which ignores the finite size of the mirror.
This is equivalent to assuming that
{\postref
the}
patterns of micro-roughness
are small in comparison to the mirror size.
We can make use of a commonly used (and empirically derived)
parameterization for the autocorrelation function,
\begin{equation}
  C(r) \, = \, \sigma^{2} \exp \left[ -(2 \pi r / \Lambda)^{2h}
  \right]  \,\, ,
  \label{eq:auto}
\end{equation}
in which $\Lambda$ is a horizontal correlation wavelength (or
coherence length scale) for the mirror and $h$ is a texture
parameter that describes the shapes of the surface features.
Small values ($h \ll 1$) indicate extremely jagged surfaces and
large values ($h \approx 1$) describe smooth crests and troughs.
{\postref
Sometimes the autocorrelation function is expressed as a sum of
several terms each similar in form to Equation (\ref{eq:auto}),
but with different values of $h$ \cite{El93}.
In that case, the r.m.s.\  surface roughness $\sigma$ in each term
can refer to different ranges of spatial scale.}
The special value $h = 1/2$ is often identified as a ``conventional
surface finish,'' and we will restrict ourselves to this value
for simplicity (see \opencite{Si88}; \opencite{CT93};
\opencite{Fn94}).
Using this value, combined with the approximation of small-scale
surface roughness ($\sigma \ll \lambda$), the intensity scattering
profile becomes
\begin{equation}
  S(\theta) \, = \, \frac{\delta(\theta)}{2\pi\theta}
  \, + \,
  \frac{8\pi \sigma^{2} \Lambda^2}{\lambda^{4} [ 1 +
  (\theta \Lambda / \lambda)^{2} ]^{3/2}} \,\, .
  \label{eq:Sinha}
\end{equation}
The first term, proportional to a Dirac delta function, describes
the ideal specular beam.
The second term is the scattering due to micro-roughness, and
its integral over all solid angles gives the so-called
total integrated scatter (TIS), or
\begin{equation}
  \mbox{TIS} \, = \,
  2\pi \int_{\theta \neq 0} \mathrm{d}\theta \, \theta \, S(\theta)
  \, = \, \frac{16 \pi^{2} \sigma^2}{\lambda^2}  \,\, .
\end{equation}
Noting that $\theta^{2} = \theta_{x}^{2} + \theta_{y}^{2}$,
we assume $\theta_{x} \neq 0$ and integrate over $\theta_y$ to
obtain
\begin{equation}
  \frac{\mathrm{d}I}{I_{0} \mathrm{d}\theta_x} \, = \,
  \frac{16 \pi \sigma^{2} \Lambda}{\lambda^{3} [ 1 +
  (\theta_{x} \Lambda / \lambda)^{2} ]} \,\, .
\end{equation}
For large angles, far from the specular beam ({\it i.e.,}
$\theta_{x} \gg \lambda / \Lambda$) this reduces to
\begin{equation}
  \frac{\mathrm{d}I}{I_{0} \mathrm{d}\theta_x} \, \approx \,
  \frac{16 \pi \sigma^{2}}{\lambda \Lambda \theta_{x}^2}
  \,\, ,
\end{equation}
which is used in Equation (\ref{eq:Sboth}) above.

\subsection{Non-Specular Diffraction due to Finite Mirror Size
and Aberrations}

For an illuminated mirror that reflects a finite-sized
(rectangular) incident beam of light, diffraction will broaden
the reflected beam and give rise to power at non-specular angles.
This situation is similar to the diffraction of parallel rays
that pass through a rectangular aperture (see, {\it e.g.,}
\opencite{Sc00}; \opencite{Hecht2002}).

The simplest case of a mirror with an ideal surface can be treated
by assuming $z = 0$ in Equation (\ref{eq:born}) and taking the
integration limits to be finite in both $x$ and $y$.
This yields
\begin{equation}
  S (\theta_{x}, \theta_{y}) \, = \, \frac{D_{x} D_{y}}{\lambda^2}
  \left( \frac{\sin u_x}{u_x} \right)^{2}
  \left( \frac{\sin u_y}{u_y} \right)^{2}
  \label{eq:DxDy}
\end{equation}
where $D_{x} = (x_{\rm max} - x_{\rm min})$ and
$D_{y} = (y_{\rm max} - y_{\rm min})$ are the dimensions of the
illuminated part of the mirror.
The dimensionless angle coordinates are defined as
$u_{i} = \pi D_{i} \theta_{i} / \lambda$, for $i = x,y$.
Equation (\ref{eq:DxDy}) is the well-known diffraction pattern
that drops off as $u^{-2}$ along the normal axes of the
rectangular aperture, and as $u^{-4}$ along the diagonals.
As above, we integrate over $\theta_y$ to obtain
\begin{equation}
  \frac{\mathrm{d}I}{I_{0} \mathrm{d}\theta_{x}} \, = \,
  \frac{D_x}{\lambda}
  \left( \frac{\sin u_x}{u_x} \right)^{2} \, \approx \,
  \frac{\lambda}{2 \pi^{2} D_{x} \theta_{x}^2}  \,\, .
  \label{eq:Sideal}
\end{equation}
The latter approximation above is taken in the limiting case of
large-angle scattering ($u_{x} \gg 1$), where the mean behavior
of the function $(\sin u / u)^2$ can be approximated as
$1 / (2u^{2})$.

For diffraction off a mirror containing large-scale aberrations,
we must again apply a variable mirror height $z(x,y)$ in
Equation (\ref{eq:born}).
This is traditionally done by inserting a phase factor that takes
account of the optical-path differences between the aberrated
wavefronts and the ideal, non-aberrated wavefronts.
\inlinecite{BW99} and \inlinecite{Sc00} defined these phase factors
in terms of Zernike polynomials for various types of common mirror
deformation patterns.
Aberrations change the diffraction profile defined above in
two distinct ways:
\begin{list}{\arabic{bean}.}{\usecounter{bean}%
\setlength{\leftmargin}{0.18in}%
\setlength{\rightmargin}{0.0in}%
\setlength{\labelwidth}{0.18in}%
\setlength{\labelsep}{0.05in}%
\setlength{\listparindent}{0.0in}%
\setlength{\itemsep}{0.05in}%
\setlength{\parsep}{0.0in}%
\setlength{\topsep}{0.05in}}
\item[{\it{i}})]
They enhance the non-specular scattering in the far wings of
the profile (see, {\it e.g.,} Figures 10.12 to 10.17 of
\opencite{Sc00}).
Here, we take this into account by multiplying
Equation (\ref{eq:Sideal}) by a wing enhancement factor $W$.
\item[{\it{ii}})]
They broaden the specular beam, which for a finite mirror or
aperture is defined according to how far from the ideal specular
ray one must go to encompass a specified level of ``encircled
energy.''
The broadening of the specular beam is related closely to how
far the actual mirror is from the diffraction limit.
\end{list}
In order to quantify these effects, we calculated a range of
diffraction profiles for various types of aberration.
\inlinecite{Ko95} noted that spherical aberrations
tended to be the dominant type for the UVCS mirrors.
The remainder of this section concentrates only on that type.
The standard measure of the magnitude of an aberration is
the r.m.s.\  wavefront error $\omega$, which is usually
expressed as a dimensionless number of wavelengths
$\omega / \lambda$.
The relative departures from the perfect (non-aberrated) case
were computed for a circular mirror, but they should be comparable
for mirrors of other shapes.

\begin{figure} 
\centerline{\includegraphics[width=0.80\textwidth,clip=]
{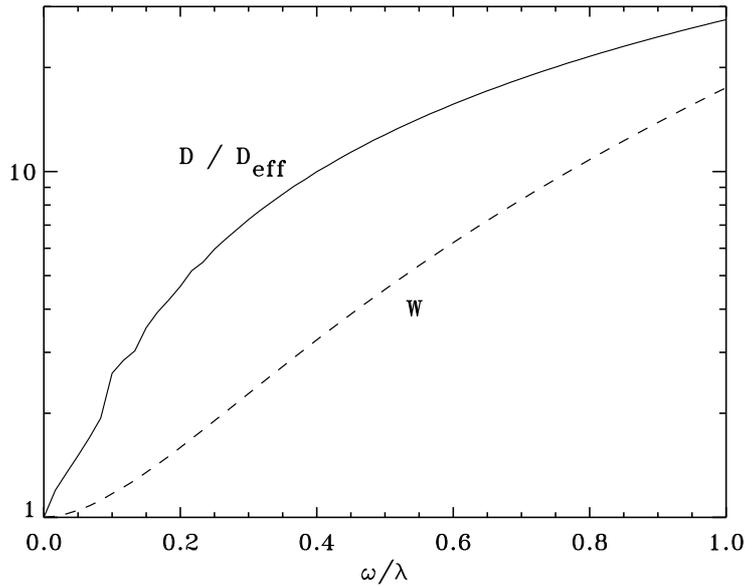}}
\caption{Effects of spherical aberrations on the ratio of the
mirror diameter $D$ to the effective diameter $D_{\rm eff}$
(solid curve) and on the relative wing enhancement $W$
(dashed curve).}
\label{fig5}
\end{figure}

Figure \ref{fig5} shows both the broadening of the specular beam
and the enhancement of the power-law scattering wing as a function
of the wavefront error.
The width of the specular beam was computed by determining the
angular radius [$\theta_s$] at which the encircled energy was 0.838
times the total energy distributed over all angles.
This fraction is equivalent to the light enclosed by the
the classical ``Airy disk'' (see \opencite{Hecht2002}).
For a diffraction-limited circular mirror without aberrations,
$\theta_{s} = 1.22 \lambda / D$.
A mirror with aberrations has a larger value of $\theta_s$
and thus corresponds to a smaller effective diameter
$D_{\rm eff}$, which would have given rise a specular beam of
identical width in the absence of aberrations.
Once $\theta_s$ has been computed for a given magnitude of the
spherical aberration, it is then possible to solve for the ratio
\begin{equation}
  \frac{D}{D_{\rm eff}} \, \approx \,
  \frac{\theta_s}{1.22 \, \lambda/D} \, \geq 1 \,\, .
  \label{eq:thetas}
\end{equation}
The quantity $D_{\rm eff}$ characterizes the
horizontal spatial scales of large wavefront deviations on
the mirror surface, just as $\Lambda$ characterizes the
coherence length of the micro-scale wavefront deviations.
Interestingly, laboratory and in-flight measurements for the UVCS
Ly$\alpha$ primary mirror yielded similar values for the ratios
$D_{x} / D_{\rm eff}$ and $D_{x} / \Lambda$ (see Section 6).

We computed the magnitude of the enhanced diffraction wing by
fitting the angular dependence of $S(\theta)$ with a power-law
(outside the specular beam).
The ratio $W$ was found by taking the ratio of the numerically
computed wing intensity to that produced by a mirror with no
aberrations.
The dependence of this ratio is also shown in Figure \ref{fig5}.
Thus, if any one of the three quantities ($\omega / \lambda$,
$D / D_{\rm eff}$, or $W$) is specified for a given mirror,
Figure \ref{fig5} can be used to estimate the other two.

\begin{acks}
This work has been supported by the National Aeronautics and Space
Administration (NASA) under grants NNX07AL72G, NNX08AQ96G,
NNX09AB27G, and NNX09AN61G to the Smithsonian Astrophysical
Observatory, by Agenzia Spaz\-i\-ale Italiana, and by the Swiss
contribution to the ESA PRODEX program.
SOHO is a project of international cooperation between
NASA and the European Space Agency (ESA).
\end{acks}

\end{article} 
\end{document}